\documentclass{aastex62}

\received{...}
\revised{...}
\accepted{...}

\submitjournal{ApJ}

\shorttitle{Failed eruption caused by reconnection}
\shortauthors{Li et al.}

\begin{document}

\title{Failure of a solar filament eruption caused by magnetic reconnection with overlying coronal loops}

\correspondingauthor{Leping Li}
\email{lepingli@nao.cas.cn}

\author[0000-0001-5776-056X]{Leping Li}
\affil{National Astronomical Observatories, Chinese Academy of Sciences, Beijing 100101, People's Republic of China}
\affiliation{Key Laboratory of Solar Activity and Space Weather, National Space Science Center, Chinese Academy of Sciences, Beijing 100190, People's Republic of China}
\affiliation{University of Chinese Academy of Sciences, Beijing 100049, People's Republic of China}

\author[0000-0001-5705-661X]{Hongqiang Song}
\affiliation{School of Space Science and Technology, Shandong University, Weihai, Shandong 264209, People's Republic of China}

\author[0000-0002-9534-1638]{Yijun Hou}
\affil{National Astronomical Observatories, Chinese Academy of Sciences, Beijing 100101, People's Republic of China}
\affiliation{Key Laboratory of Solar Activity and Space Weather, National Space Science Center, Chinese Academy of Sciences, Beijing 100190, People's Republic of China}
\affiliation{University of Chinese Academy of Sciences, Beijing 100049, People's Republic of China}
\affiliation{Yunnan Key Laboratory of Solar Physics and Space Science, Kunming 650216, People's Republic of China}

\author[0000-0001-8228-565X]{Guiping Zhou}
\author[0000-0003-2047-9664]{Baolin Tan}
\affil{National Astronomical Observatories, Chinese Academy of Sciences, Beijing 100101, People's Republic of China}
\affiliation{Key Laboratory of Solar Activity and Space Weather, National Space Science Center, Chinese Academy of Sciences, Beijing 100190, People's Republic of China}
\affiliation{University of Chinese Academy of Sciences, Beijing 100049, People's Republic of China}

\author[0000-0001-8950-3875]{Kaifan Ji}
\author[0000-0002-5261-6523]{Yongyuan Xiang}
\affiliation{Yunnan Observatories, Chinese Academy of Sciences, Kunming 650216, People's Republic of China}

\author[0000-0003-4804-5673]{Zhenyong Hou}
\affiliation{School of Earth and Space Sciences, Peking University, Beijing 100871, People's Republic of China}

\author[0000-0002-9293-8439]{Yang Guo}
\affiliation{School of Astronomy and Space Science and Key Laboratory of Modern Astronomy and Astrophysics, Nanjing University, Nanjing 210023, People's Republic of China}

\author[0000-0002-1190-0173]{Ye Qiu}
\affiliation{Institute of Science and Technology for Deep Space Exploration, Suzhou Campus, Nanjing University, Suzhou 215163, People's Republic of China}

\author[0000-0001-9647-2149]{Yingna Su}
\author[0000-0002-5898-2284]{Haisheng Ji}
\affiliation{Key Laboratory of Dark Matter and Space Astronomy, Purple Mountain Observatory, Chinese Academy of Sciences, Nanjing 210023, People's Republic of China}
\affiliation{School of Astronomy and Space Science, University of Science and Technology of China, Hefei, Anhui 230026, People's Republic of China}

\author[0000-0003-4078-2265]{Qingmin Zhang}
\affiliation{Key Laboratory of Dark Matter and Space Astronomy, Purple Mountain Observatory, Chinese Academy of Sciences, Nanjing 210023, People's Republic of China}
\affiliation{Yunnan Key Laboratory of Solar Physics and Space Science, Kunming 650216, People's Republic of China}

\author{Yudi Ou}
\affiliation{Key Laboratory of Dark Matter and Space Astronomy, Purple Mountain Observatory, Chinese Academy of Sciences, Nanjing 210023, People's Republic of China}
\affiliation{School of Astronomy and Space Science, University of Science and Technology of China, Hefei, Anhui 230026, People's Republic of China}

\begin{abstract}

Failure of a filament eruption caused by magnetic reconnection between the erupting filament and the overlying magnetic field has been previously proposed in numerical simulations.
It is, however, rarely observed.
In this study, we report the reconnection between an erupting filament and its overlying coronal loops, that results in the failure of the filament eruption.
On 2023 September 24, a filament was located in active region 13445.
It slowly rose, quickly erupted, rapidly decelerated, and finally stopped, with an untwisting motion.
As a failed eruption, the event is associated with an M4.4 flare but no coronal mass ejection.
During the eruption, the filament became bright, and the overlying loops appeared first in the high-temperature channels.
They have average temperatures of $\sim$12.8 and $\sim$9.6\,MK, respectively, indicating that both of them were heated.
Two sets of new loops, separately connecting the filament endpoints and the overlying loop footpoints, then formed.
Subsequently, the heated overlying loops were seen sequentially in the low-temperature channels, showing the cooling process, which is also supported by the light curves.
Plasmoids formed, and propagated bidirectionally along the filament and the overlying loops, indicating the presence of plasmoid instability.
These results suggest that reconnection occurs between the erupting filament and the overlying loops.
The erupting filament eventually disappeared, with the appearance of more newly-formed loops.
We propose that the reconnection between the erupting filament and the overlying loops ruins the filament completely, and hence results in the failed eruption.

\end{abstract}


\section{Introduction} \label{sec:int}

Solar filaments (prominences), composed of cooler, denser chromospheric material suspended in the hotter, rarer corona, are elongated features along magnetic polarity-inversion lines \citep[PILs;][]{1991A&A...252..353S, 2010SSRv..151..333M, 2013SoPh..282..147L, 2014LRSP...11....1P}.
They are seen as dark filaments on the disk and bright prominences above the limb in the H$\alpha$ and extreme ultraviolet (EUV) passbands \citep{2015ApJS..219...17Y, 2016ApJ...816...41Y, 2021ApJ...919L..21L}.
Filaments sometimes erupt \citep{2001ApJ...554..474Z, 2007ChJAA...7..129J, 2015ApJ...804L..38S, 2016NatCo...711837X, 2023ApJ...949...66L}.
Based on the amount of ejected mass and magnetic structures, filament eruptions have been classified into three types: full, partial, and failed (or confined) eruptions \citep{2007SoPh..245..287G}. 
Among them, the first two are associated with flares and coronal mass ejections \citep[CMEs;][]{2001JGR...10625053L, 2011LRSP....8....1C, 2012ApJ...750...12S, 2023ApJ...959...69H}, the major driver of severe space weather, while the last one, that decelerates and stops at a certain height after an initial acceleration, has no accompanying CME \citep{2003ApJ...595L.135J, 2022ApJ...932L...9K, 2022ApJ...941L...1L, 2024MNRAS.530..473X}.

For the failed eruption, a prevailing model is that a magnetic flux rope (MFR), a set of magnetic field lines winding around a common axis, runs into kink instability but does not reach the threshold height for torus instability, which is determined by a critical decay index of the overlying field \citep{2005ApJ...630L..97T, 2006PhRvL..96y5002K, 2016ApJ...832..106H}.
Here, the decay index describes how fast the external poloidal field declines with height.
The model is then observationally supported by many case and statistical studies \citep[e.g.,][]{2007ApJ...665.1428W, 2010ApJ...725L..38G, 2011ApJ...732...87C, 2013ApJ...778...70C, 2015ApJ...804L..28S}.
However, \citet{2019ApJ...877L..28Z} noticed that several filament eruptions, where decay indexes of external field exceed the theoretical expectation of the torus instability, do not result into CMEs.
They suggested that the occurrence of torus instability is not a sufficient condition for the successful eruption, and  the rotation motion of the filament has a certain correlation with the failed eruption \citep{2018ApJ...864L..37S}.
Recently, performing a data-driven magnetohydrodynamic (MHD) simulation, \citet{2021NatCo..12.2734Z} proposed that the Lorentz force component, induced by the non-axisymmetry of MFR, plays a major role in causing the failed eruption. 

In the failed eruption, magnetic reconnection between the erupting filament and the overlying field is also suggested to play a vital role \citep{1999ApJ...515L..81A, 2016ApJ...832..106H}.
Reconstructing the configuration of an intermediate filament, \citet{2022ApJ...928..160Z} analyzed how it is first triggered to erupt and then fails to escape using MHD simulations, and found that reconnection between the erupting MFR and the overlying field is the key factor constraining the filament eruption.
Recently, \citet{2023MNRAS.525.5857J} presented a new model of failed eruption where reconnection plays a leading role in the initiation and failure of the eruption.
In their model, during the rise, the MFR experiences a significant rotation around the vertical axis, that renders the field direction at the MFR apex almost inverse to the overlying field.
A strong current sheet is then formed between the MFR and the overlying flux.
Reconnection occurs in the current sheet, ruins the MFR completely, and results in the failed eruption.
In the specific topology of a quadrupolar active region (AR), \citet{2023ApJ...951L..35C} modeled an MFR eruption, and proposed that external reconnection between the erupting MFR and its overlying flux can play a vital role in confining the eruption.

In observations, reconnection between the erupting filament and the nearby coronal loops, leading to the failed eruption, has been previously reported \citep[e.g.,][]{li16a, 2017ApJ...843...93C}.
Analyzing three failed filament eruptions, \citet{2012A&A...548A..89N} presented EUV brightenings and in one case very weak hard X-ray emission, produced by the interaction between the erupting filament and the overlying loops, and regarded them as observational evidence of reconnection. 
In a failed filament eruption, \citet{2017ApJ...838...15L} found brightenings in the helical structures, and suggested that reconnection takes place between the expanding sheared filament arcade and the surrounding loops.
Recently, \citet{2023ApJ...959...67C} identified external reconnection signatures in a quadrupolar configuration during a failed filament eruption, and proposed that the external reconnection between the erupting MFR and the overlying loops results in the failed eruption. 

During the failed eruption, reconnection between the erupting filament and the overlying field has been previously presented in both the simulations and observations \citep{1999ApJ...515L..81A, 2012A&A...548A..89N, 2016ApJ...832..106H, 2017ApJ...838...15L, 2021ApJ...923..142C, 2023ApJ...951L..35C, 2023ApJ...959...67C, 2023MNRAS.525.5857J}.
Details of the reconnection are, however, rarely reported.
In this study, we present the detailed evolution of reconnection between an erupting filament and its overlying loops, that results in the failure of the filament eruption.
The observations, results, and summary and discussion are illustrated in Sections\,\ref{sec:obs}-\ref{sec:sum}, respectively.

\section{Observations}\label{sec:obs}

The Atmospheric Imaging Assembly \citep[AIA;][]{2012SoPh..275...17L} on board the Solar Dynamics Observatory \citep[SDO;][]{2012SoPh..275....3P} is a set of normal-incidence imaging telescopes.
It acquires solar atmospheric images in 10 wavelength passbands.
Different AIA passbands show plasma at different temperatures, e.g., 131\,\AA~peaks at $\sim$10\,MK (Fe\,{\sc{xxi}}) and $\sim$0.6\,MK (Fe\,{\sc{viii}}), 94\,\AA~peaks at $\sim$7.2\,MK (Fe\,{\sc{xvii}}), 335\,\AA~peaks at $\sim$2.5\,MK (Fe\,{\sc{xvi}}), 211\,\AA~peaks at $\sim$1.9\,MK (Fe\,{\sc{xiv}}), 193\,\AA~peaks at $\sim$1.5\,MK (Fe\,{\sc{xii}}), 171\,\AA~peaks at $\sim$0.9\,MK (Fe\,{\sc{ix}}), and 304\,\AA~peaks at $\sim$0.05\,MK (He\,{\sc{ii}}).
The spatial sampling and time cadence of AIA EUV images are 0.6\arcsec\,pixel$^{-1}$ and 12\,s, respectively.
The Helioseismic and Magnetic Imager \citep[HMI;][]{2012SoPh..275..229S} on board the SDO provides line-of-sight (LOS) magnetograms, with a time cadence and spatial sampling of 45\,s and 0.5\arcsec\,pixel$^{-1}$, respectively.

The New Vacuum Solar Telescope \citep[NVST;][]{2014RAA....14..705L} is a 1-m ground-based solar telescope, located in the Fuxian Solar Observatory of the Yunnan Observatories, Chinese Academy of Sciences (CAS).
It provides observations of the solar fine structures and their evolution in the solar lower atmosphere.
On 2023 September 24, the NVST observed AR 13445 from 03:11 to 06:27\,UT, with a field-of-view (FOV) of 185\arcsec$\times$185\arcsec~in the H$\alpha$ channel, centered at 6562.8\,\AA~with a bandwidth of 0.25\,\AA.
The spatial sampling and time cadence of the NVST H$\alpha$ images are 0.168\arcsec\,pixel$^{-1}$ and 41.3\,s, respectively.

The Chinese H$\alpha$ Solar Explorer \citep[CHASE;][]{2022SCPMA..6589602L}, lauched on 2021 October 14, is the first solar space mission of the China National Space Administration (CNSA).
The H$\alpha$ Imaging Spectrograph \citep[HIS;][]{2022SCPMA..6589605L} is the scientific payload of CHASE.
It can provide spectroscopic observations of the Sun by scanning the full solar disk in both the H$\alpha$ (6559.7-6565.9\,\AA) and Fe\,{\sc{i}} (6567.8-6570.6\,\AA) passbands.
The data of CHASE/HIS has been calibrated through dark-field, flat-field, and slit image curvature corrections, wavelength and intensity calibration, and coordinate transformation \citep{2022SCPMA..6589603Q}.
The time cadence and spatial sampling of HIS H$\alpha$ images are $\sim$71\,s and 1.05\arcsec\,pixel$^{-1}$, respectively.

The Solar Upper Transition Region Imager \citep[SUTRI;][]{2023RAA....23f5014B} on board the Space Advanced Technology demonstration satellite (SATech-01) was launched into a Sun-synchronous orbit at a height of $\sim$500\,km by the CAS in 2022 July.
It aims to establish connections between structures in the lower solar atmosphere and corona, and advance our understanding of various types of solar activity, such as flares, filament eruptions, and CMEs.
SUTRI takes full-disk solar atmospheric images, with a FOV of $\sim$41.6\arcmin$\times$41.6\arcmin~and moderate spatial sampling of $\sim$1.23\arcsec\,pixel$^{-1}$, at the Ne\,{\sc{vii}} 465\,\AA~spectral line, formed at the characteristic temperature of $\sim$0.5\,MK, similar to the lower characteristic temperature ($\sim$0.6\,MK) of the AIA 131\,\AA~channel, with a filter width of $\sim$30\,\AA~ \citep{2017RAA....17..110T}.
As SATech-01 is not a solar-dedicated spacecraft, the solar observation time is $\sim$16\,hr each day because the Earth eclipse time accounts for $\sim$1/3 of SATech-01's orbital period, and the normal time cadence of SUTRI images is $\sim$30\,s.

In this study, we employ the AIA 131, 94, 335, 211, 193, 171, and 304\,\AA, SUTRI 465\,\AA, and NVST and CHASE H$\alpha$ images to investigate the filament eruption and the reconnection between the erupting filament and the overlying loops on 2023 September 24.
To better show the evolution, the AIA, SUTRI, and CHASE images are enhanced by using the multiscale Gaussian normalization (MGN) technique \citep{2014SoPh..289.2945M}.
HMI LOS magnetograms are used to analyze the evolution of magnetic fields associated with the filament eruption and reconnection.
GOES-16	soft X-ray 1-8\,\AA~flux with a time cadence of 1\,s is employed to study the flare associated with the filament eruption.
All the data from different instruments, i.e., the SDO, SUTRI, NVST, and CHASE, and passbands have been aligned with AIA 304\,\AA~images by using an automatic mapping technique \citep{ji19}. 
Details of the data analyzed in this paper are listed in Table\,\ref{tab:data}. 

\section{Results}\label{sec:res}

\subsection{Filament eruption}\label{sec:eruption}

\begin{figure}[ht!]
\centering
\includegraphics[width=0.8\textwidth]{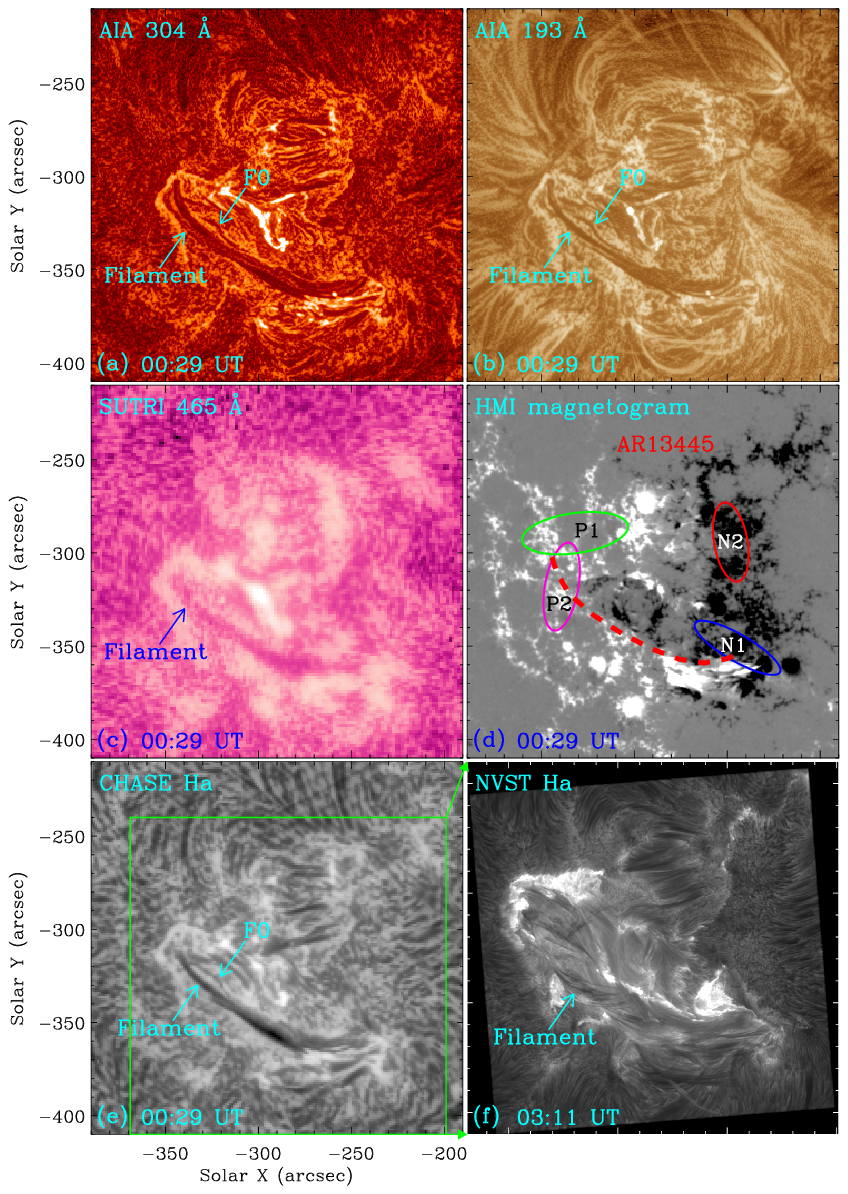}
\caption{General information of the filament.
(a) SDO/AIA 304 and (b) 193\,\AA~images, (c) SUTRI 465\,\AA~image, (d) SDO/HMI LOS magnetogram, and (e) CHASE and (f) NVST H$\alpha$ images.
The AIA, SUTRI, and CHASE images are enhanced by the MGN technique.
F0 in (a), (b), and (e) represents another filament underneath the erupting filament.
In (d), the red dashed line indicates the filament in (a), ending in the positive and negative magnetic fields, P1 and N1, enclosed by the green and blue ellipses (see Figure\,\ref{f:reconnection}(b)), and the pink and red ellipses enclose the positive and negative  fields, P2 and N2, connected by the loops L3 overlying the filament (see Figure\,\ref{f:reconnection}(n)).
The green rectangle in (e) shows the FOV of (f).
See Section\,\ref{sec:res} for details.
\label{f:general}}
\end{figure}

On 2023 September 24, AR 13445 was located at the heliographic position S21\,E22, with main positive and negative magnetic fields in the east and west, respectively (see Figure\,\ref{f:general}(d)). 
In the center and the south of the AR, the interlaced positive and negative fields are identified (see Figure\,\ref{f:general}(d)).
A filament, with a length and width of $\sim$98 and $\sim$6\,Mm, was observed in the AR (see Figures\,\ref{f:general}(a) and (b)).
It ends in the positive and negative fields, P1 and N1, enclosed by the green and blue ellipses in Figure\,\ref{f:general}(d), and is located along the southern PILs of the AR (see the red dashed line in Figure\,\ref{f:general}(d)).
The filament was also observed by the SUTRI 465\,\AA, and CHASE and NVST H$\alpha$ images (see Figures\,\ref{f:general}(c), (e), and (f)).
Underneath the filament which is about to erupt, another filament F0 is identified (see Figures\,\ref{f:general}(a), (b), and (e)), and will be clearly seen and remain stationary after the filament eruption  (see Figure\,\ref{f:reconnection}(l)).
These two filaments form a double-decker structure.

From $\sim$02:48\,UT, a brightening took place in the northeastern part of the filament (see Figure\,\ref{f:reconnection}(a)), and propagated bidirectionally along the filament (see the online animated version of Figure\,\ref{f:reconnection}).
The filament then erupted (see Figure\,\ref{f:reconnection}(b)).
During the eruption, an untwisting motion toward the northern endpoint of the filament is identified (see Figures\,\ref{f:reconnection}(b) and (c)).
Subsequently, more twisted threads of the filament disappeared, while less twisted threads became visible.
Moreover, both endpoints of the filament, enclosed by the green and blue ellipses in Figure\,\ref{f:reconnection}(b), expanded.
Additionally, the northern endpoint changed in time from east to west in the green ellipse in Figure\,\ref{f:reconnection}(b) (see the online animated version of Figure\,\ref{f:reconnection}).
The eruption and untwisting motion of the filament are also recorded by the NVST and CHASE H$\alpha$ and SUTRI 465\,\AA~images (see Figure\,\ref{f:general}(f) and the online animated version of Figure\,\ref{f:reconnection}).

Along the erupting direction, e.g., the AB direction, in the cyan rectangle in Figure\,\ref{f:reconnection}(b), a time slice of AIA 304\,\AA~images is made, as displayed in Figure\,\ref{f:measurements}(a).
The filament rose slowly from $\sim$02:50\,UT after the appearance of the brightening, and then erupted quickly with a mean projection acceleration and speed of 29\,m\,s$^{-2}$ and 45\,km\,s$^{-1}$, respectively (see the cyan dotted line in Figure\,\ref{f:measurements}(a)).
Subsequently, it decelerated rapidly, and finally stopped erupting (see Figure\,\ref{f:measurements}(a)).
Moreover, the untwisting motion of the filament is also clearly detected.

The GOES-16 soft X-ray 1-8\,\AA~flux increased from $\sim$02:51\,UT, immediately after the beginning of the slow rise of the filament (see Figure\,\ref{f:lightcurves}(a)).
It peaked at 03:28\,UT (see the red vertical dotted line in Figure\,\ref{f:lightcurves}(a)), and then decreased slowly.
An M4.4 class flare took place associated with the filament eruption.
However, no CME, associated with the filament eruption, is observed in the white-light coronagraphs.
This further supports that the filament eruption is a failed eruption.

\subsection{Magnetic reconnection between the erupting filament and the overlying loops}\label{sec:reconnection}

\begin{figure}[ht!]
\centering
\includegraphics[width=1.\textwidth]{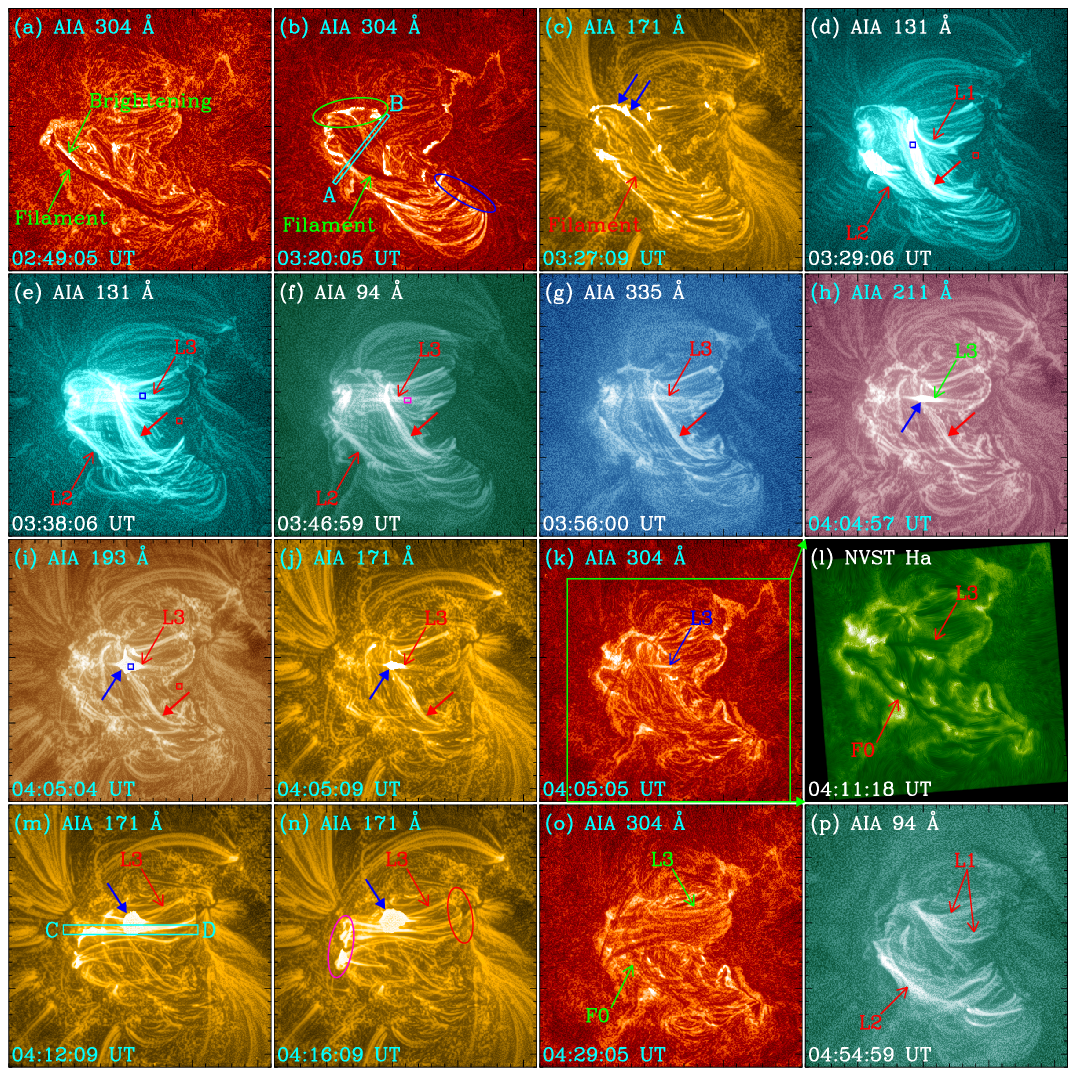}
\caption{Magnetic reconnection between the erupting filament and the overlying loops.
(a), (b), (k), (o) AIA 304, (c), (j), (m), (n) 171, (d), (e) 131, (f), (p) 94, (g) 335, (h) 211, and (i) 193\,\AA, and (l) NVST Ha images.
The AIA images are enhanced by the MGN technique.
The cyan rectangles AB and CD in (b) and (m) show the positions for the time slices in Figures\,\ref{f:measurements}(a) and (b), respectively.
The green and blue ellipses in (b) enclose the filament endpoints.
The blue solid arrows in (c), (h)-(j), (m), and (n) denote the plasmoids.
The blue and red rectangles in (d), (e), and (i) separately enclose the regions for the DEM curves in Figures\,\ref{f:measurements}(c)-(e) and the locations where the background emission is calculated.
The red solid arrows in (d)-(j) mark the filament.
The pink rectangle in (f) marks the region for the light curves in Figures\,\ref{f:lightcurves}(b)-(i).
The green rectangle in (k) shows the FOV of (l).
F0 in (l) and (o) represents the filament underneath the erupting filament.
The pink and red ellipses in (n) enclose the footpoints of overlying loops L3.
An animation of the unannotated AIA and NVST images (panels (e)-(l)) is available.
It covers $\sim$3.5\,hr starting at 02:30\,UT, with a time cadence of 1\,minute.
The FOVs of (a)-(k) and (m)-(p), and (l) are the same as those of Figures\,\ref{f:general}(a)-(e) and (f), respectively.
See Section\,\ref{sec:res} for details.
(An animation of this figure is available.) 
\label{f:reconnection}}
\end{figure}

During the filament eruption, bright plasmoids, marked by the blue solid arrows in Figure\,\ref{f:reconnection}(c), formed in the northern part of the filament, and propagated bidirectionally along the filament (see the online animated version of Figure\,\ref{f:reconnection}).
Part of the erupting filament then became bright, denoted by the red solid arrows in Figures\,\ref{f:reconnection}(d)-(j).
Using six AIA EUV passbands, including 131, 94, 335, 211, 193, and 171\,\AA, we analyze the temperature and emission measure (EM) of the bright filament.
Here, we employ the differential EM (DEM) analysis using ``xrt\_dem\_iterative2.pro" \citep{2004IAUS..223..321W, 2012ApJ...761...62C}.
In order to estimate the DEM uncertainties, 100 Monte Carlo realizations of the data are computed.
The filament region, enclosed by the blue rectangle in Figure\,\ref{f:reconnection}(d), is chosen to compute the DEM.
The nearby quiet region, enclosed by the red rectangle in Figure\,\ref{f:reconnection}(d), outside the filament is selected for the background emission that is subtracted from the filament region.
In each region, the digital number counts in each of the six AIA EUV channels are temporally normalized by the exposure time and spatially averaged over all pixels.

\begin{figure}[ht!]
\centering
\includegraphics[width=0.66\textwidth]{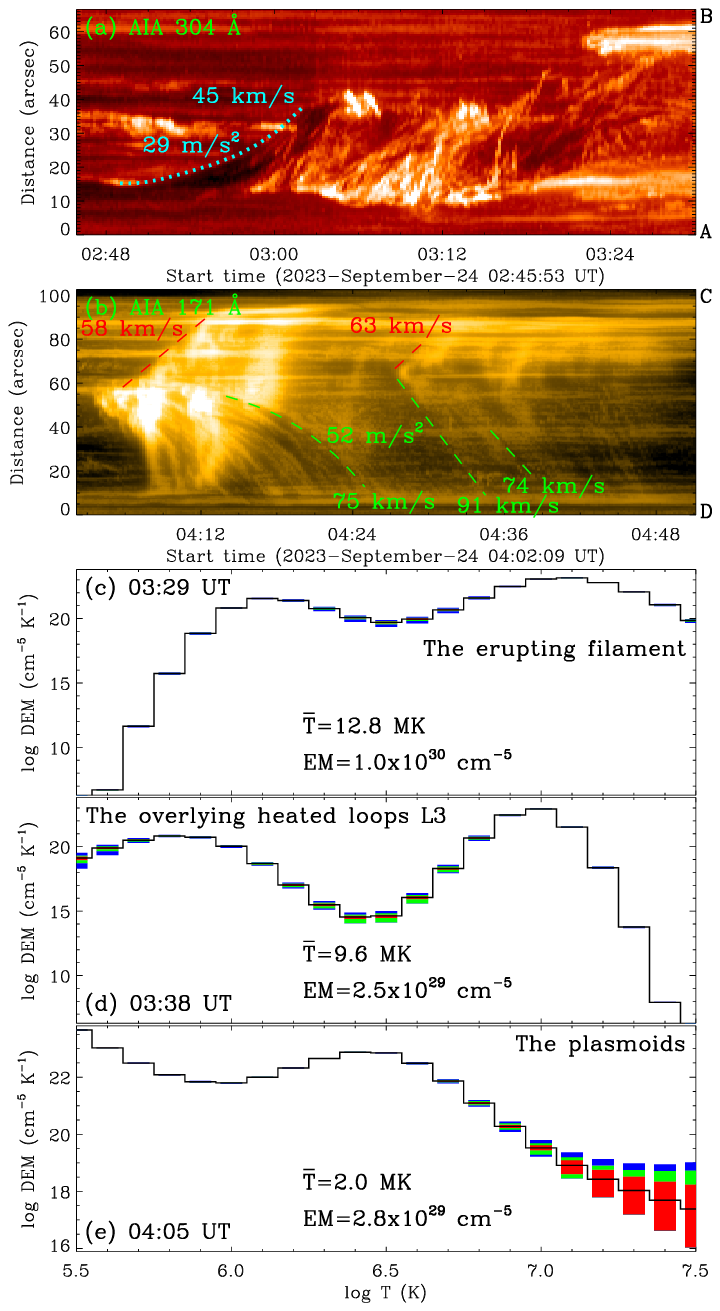}
\caption{Evolution of the filament eruption and magnetic reconnection.
(a), (b) Time slices of AIA 304 and 171\,\AA~images along the AB and CD directions in the cyan rectangles in Figures\,\ref{f:reconnection}(b) and (m), and (c)-(e) DEM curves for the regions enclosed by the blue rectangles in Figures\,\ref{f:reconnection}(d), (e), and (i) of the erupting filament, the overlying heated loops L3, and the plasmoids, respectively.
The cyan dotted line in (a) outlines the filament eruption, and the red and green dashed lines in (b) show the plasma motions.
In (c)-(e), the black curves are the best-fit DEM distributions, and the red, green, and blue rectangles represent the regions containing 50\%, 51-80\%, and 81-95\% of the Monte Carlo solutions, respectively.
See Section\,\ref{sec:res} for details.
\label{f:measurements}}
\end{figure}

The DEM curve of the filament region at 03:29\,UT is shown in Figure\,\ref{f:measurements}(c).
It is well constrained with small errors.
The main peak of the DEM curve lies at 12.6\,MK (6.5$\le\log T (K)\le$7.5), and the secondary peak, 2 orders of magnitude smaller than the main peak, appears at $\sim$1.3\,MK (5.5$\le\log T (K)\le$6.5). 
The filament thus became a multithermal structure during the eruption.
The DEM-weighted average temperature and EM, over the temperature range of 5.5$\le\log T (K)\le$7.5, are calculated to be 12.8\,MK and 1.0$\times$10$^{30}$\,cm$^{-5}$, respectively.
This indicates that hot plasma dominates the emission in the filament region.

Employing the EM, the electron number density ($n_{e}$) of the filament is estimated using 
$n_{e}$=$\sqrt{\frac{EM}{D}}$, where $D$ is the LOS depth of the filament.
Assuming that the depth ($D$) equals the width ($W$) of the filament, then the electron number density is $n_{e}$=$\sqrt{\frac{EM}{W}}$.
We measure the filament width at the place and time at which we calculate the DEM.
First, we get the intensity profile perpendicular to the filament in the AIA 131\,\AA~passband.
Using the mean intensity surrounding the filament as the background emission, we subtract it from the intensity profile.
Fitting the residual intensity profile using a single Gaussian, we obtain the FWHM of the single Gaussian fit as the width.
The filament width is measured to be $\sim$5.9\,Mm.
Employing EM=1.0$\times$10$^{30}$\,cm$^{-5}$ and W=5.9\,Mm, the electron number density is then estimated to be 4.1$\times$10$^{10}$\,cm$^{-3}$.

The loops L3, overlying the northern part of the filament, appeared first in the AIA 131\,\AA~channel (see Figure\,\ref{f:reconnection}(e)).
They connect the positive and negative fields, P2 and N2, enclosed by the pink and red ellipses in Figures\,\ref{f:general}(d) and \ref{f:reconnection}(n).
The loops are not observed at the same time, and will be seen during the subsequent cooling process in the AIA low-temperature, e.g., 335, 211, 193, 171, and 304\,\AA, and SUTRI 465\,\AA~channels.
They are hence heated up to $\sim$10\,MK, the higher characteristic temperature of the 131\,\AA~channel.
The loop region, enclosed by the blue rectangle in Figure\,\ref{f:reconnection}(e), is selected to compute the DEM.
The nearby quiet region outside the loops, enclosed by the red rectangle in Figure\,\ref{f:reconnection}(e), is chosen for the background emission that is subtracted from the loop region.

The DEM curve of the loop region at 03:38\,UT is displayed in Figure\,\ref{f:measurements}(d).
It is well constrained with small errors.
The main peak of the DEM curve appears at 10\,MK (6.5$\le\log T (K)\le$7.3), and the secondary peak, 2 orders of magnitude smaller than the main peak, lies at $\sim$0.6\,MK (5.5$\le\log T (K)\le$6.4). 
Between them, the high-temperature plasma corresponds to the loops L3, and the low-temperature plasma originates from the underlying warm loops that are superimposed on the loops L3 along the LOS (see the online animated version of Figure\,\ref{f:reconnection}).
We compute the DEM-weighted average temperature and EM over the temperature range of 5.5$\le\log T (K)\le$7.5, and get values of 9.6\,MK and 2.5$\times$10$^{29}$\,cm$^{-5}$, respectively.
This supports that hot plasma dominates the emission in the loop region.
The loops L3 were thus heated up to $\sim$9.6\,MK, consistent with the AIA imaging observations that the loops appear only in AIA high-temperature, rather than low-temperature, images.
We measure the loop width at the place and time at which we calculate the DEM, and obtain a value of $\sim$3.2\,Mm.
Employing EM=2.5$\times$10$^{29}$\,cm$^{-5}$ and W=3.2\,Mm, the electron number density of the loops is estimated to be 2.8$\times$10$^{10}$\,cm$^{-3}$, smaller than that of the erupting filament.

\begin{figure}[ht!]
\centering
\includegraphics[width=0.7\textwidth]{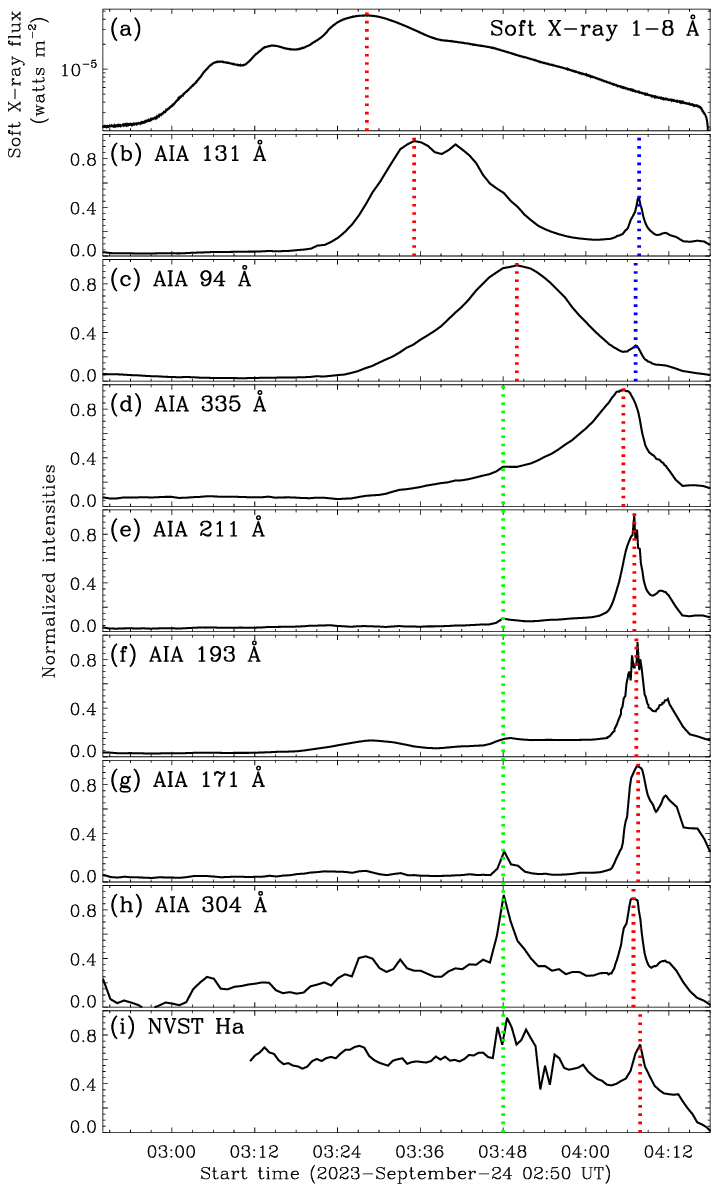}
\caption{Light curves of the overlying loops.
(a) GOES-16 soft X-ray 1-8\,\AA~flux, and (b)-(i) light curves of the AIA 131, 94, 335, 211, 193, 171, and 304\,\AA,~and NVST H$\alpha$ channels in the pink rectangle in Figure\,\ref{f:reconnection}(f).
The red, blue, and green vertical dotted lines mark the peaks of light curves. 
See Section\,\ref{sec:res} for details.
\label{f:lightcurves}}
\end{figure}

The loops L3 then appeared sequentially in AIA 94, 335, 211, 193, 171, and 304\,\AA~images (see Figures\,\ref{f:reconnection}(f)-(k) and the online animated version of Figure\,\ref{f:reconnection}), showing the cooling process of the heated loops.
In the loop region, enclosed by the pink rectangle in Figure\,\ref{f:reconnection}(f), we measure the light curves of the AIA 131, 94, 335, 211, 193, 171, and 304\,\AA~channels, and display them in Figures\,\ref{f:lightcurves}(b)-(h).
The light curves increase sequentially, peak separately at 03:35:06 (131\,\AA), 03:49:59 (94\,\AA), 04:05:24 (335\,\AA), 04:07:00 (211\,\AA), 04:07:19 (193\,\AA), 04:07:33 (171\,\AA), and 04:06:53 (304\,\AA) UT (see the red vertical dotted lines in Figures\,\ref{f:lightcurves}(b)-(h)), later than the peak (03:28\,UT) of the GOES-16 soft X-ray 1-8\,\AA~flux, and then decrease.
This shows the sequential appearance of loops L3 in AIA 131, 94, 335, 211, 193, 171, and 304\,\AA~images.
Here, the peaks of the AIA 335, 211, 193, 171, and 304\,\AA~light curves at $\sim$03:48\,UT (see the green vertical dotted lines in Figures\,\ref{f:lightcurves}(d)-(h)) come from an underlying brightening, rather than the loops L3.
Moreover, smaller peaks of the AIA 131 and 94\,\AA~light curves at 04:07:42 and 04:07:11 UT, respectively, are detected (see the blue vertical dotted lines in Figures\,\ref{f:lightcurves}(b) and (c)), representing the emission from plasma of loops L3 with the lower characteristic temperatures of the AIA 131 and 94\,\AA~channels.
These results further support the cooling process of the heated loops L3.

Plasmoids formed at the middle (top) of loops L3, denoted by the blue solid arrows in Figures\,\ref{f:reconnection}(h)-(j), (m), and (n), and then propagated along the loops L3 bidirectionally.
They are clearly observed in the AIA low-temperature channels, e.g., 211, 193, and 171\,\AA~(see the online animated version of Figure\,\ref{f:reconnection}).
As denoted by the blue rectangle in Figure\,\ref{f:reconnection}(i), we calculate the DEM of the plasmoid region at 04:05\,UT, and show it in Figure\,\ref{f:measurements}(e).
Here, the background emission, which is subtracted from the plasmoid region, is computed in the nearby quiet region, enclosed by the red rectangle in Figure\,\ref{f:reconnection}(i).
Compared with the DEM of the filament and the loops L3 in Figures\,\ref{f:measurements}(c) and (d), the plasmoid emission is well constrained at lower temperatures (5.5$\le\log T (K)\le$7.0), rather than the high temperatures, due to the weak signal in the AIA high-temperature channels, e.g., 131 and 94\,\AA.
Over the temperature range of 5.5$\le\log T (K)\le$7.0, we calculate the DEM-weighted average temperature and EM, and get values of 2.0\,MK and 2.8$\times$10$^{29}$\,cm$^{-5}$.
This supports that warm plasma dominates the emission in the plasmoid region.
Using AIA 171\,\AA~images, we measure the width of the plasmoid at the place and time at which we compute the DEM, and obtain a value of $\sim$2.5\,Mm.
Employing the EM and width, we estimate the electron number density of the plasmoid, and get a value of 3.3$\times$10$^{10}$\,cm$^{-3}$, larger (smaller) than that of the heated loops (filament).

The plasmoids propagated bidirectionally along the loops L3 toward the loop footpoints (see Figure\,\ref{f:reconnection}(m) and the online animated version of Figure\,\ref{f:reconnection}).
Dark material then quickly appeared at the middle (top) of loops L3, and moved bidirectionally toward the loop footpoints (see Figure\,\ref{f:reconnection}(o)).
Along the CD direction in the cyan rectangle in Figure\,\ref{f:reconnection}(m), we make a time slice of AIA 171\,\AA~images, and show it in Figure\,\ref{f:measurements}(b).
Evident flows, forming at the middle (top) of loops L3 and moving bidirectionally along loops L3, are clearly detected (see the red and green dashed lines in Figure\,\ref{f:measurements}(b)), with mean speeds and accelerations of $\sim$50-100\,km\,s$^{-1}$ and 52\,m\,s$^{-2}$.

The appearance of loops L3 and the material moving along the loops are also observed in SUTRI 465\,\AA~and NVST H$\alpha$ images (see Figure\,\ref{f:reconnection}(l)).
The light curve of the NVST H$\alpha$ channel is also calculated, and displayed in Figure\,\ref{f:lightcurves}(i).
It evolves with time in a manner consistent with the AIA 304\,\AA~light curve (see Figures\,\ref{f:lightcurves}(h) and (i)).
A peak at 04:07:51\,UT (see the red vertical dotted line in Figure\,\ref{f:lightcurves}(i)), similar to those of the AIA 211, 193, 171, and 304\,\AA~light curves, is identified.
Here, the AIA 211, 193, 171, and 304\,\AA, and NVST H$\alpha$ light curves peak not exactly  in the sequence of the decreasing characteristic temperatures of these channels.
This may be caused by the fact that these light curves are contaminated by the dark material that dominates the loops L3 in AIA 304\,\AA~and NVST H$\alpha$ images (see Figures\,\ref{f:reconnection}(l) and (o)).
Moreover, the peak at $\sim$03:48\,UT (see the green dotted line in Figure\,\ref{f:lightcurves}(i)), originating from the underlying brightening, is also detected.

During the filament eruption, two sets of new loops, L1 and L2, formed (see Figures\,\ref{f:reconnection}(d) and (e)).
The newly-formed loops L1 connect the positive and negative fields, P1 and N2, where the northern endpoint of filament and the western footpoint of loops L3 root, while the newly-formed loops L2 link the positive and negative fields, P2 and N1, where the eastern footpoint of loops L3 and the southern endpoint of filament root (see Figures\,\ref{f:general}(d) and \ref{f:reconnection}(b) and (n)).
The erupting filament finally disappeared, and more newly-formed loops L1 and L2 appeared (see Figure\,\ref{f:reconnection}(p) and the online animated version of Figure\,\ref{f:reconnection}).

All the results, including the heating of erupting filament and its overlying loops L3, the formation and propagation of plasmoids along the filament and the loops L3, the disappearance of erupting filament, and the appearance of newly-formed loops L1 and L2, indicate that reconnection between the erupting filament and its overlying loops L3 takes place, with the formation of newly-reconnected loops L1 and L2.
After the eruption, underneath the erupting filament, the preexisting filament F0 remained stationary (see Figures\,\ref{f:reconnection}(l) and (o)), showing that the eruption is a partial eruption of the double-decker filament.

\section{Summary and discussion}\label{sec:sum}

Employing AIA, SUTRI, CHASE, and NVST images, HMI LOS magnetograms, and GOES-16 soft X-ray flux, we investigate the failed, partial filament eruption in AR 13445 on 2023 September 24.
A brightening occurred in the northern part of the filament, and moved bidirectionally along the filament.
The filament then slowly rose, and quickly erupted, showing the untwisting motion toward the northern endpoint, the expansion of the endpoints, and the shift of the northern endpoint.
Subsequently, it decelerated, and finally stopped, indicating the failed eruption.
An M4.4 flare and no CME, associated with the failed eruption, are observed.
During the eruption, the filament became bright, with a DEM-weighted average temperature of 12.8\,MK, and the loops L3 overlying the filament simultaneously appeared first in AIA 131\,\AA~images, with a DEM-weighted average temperature of 9.6\,MK.
This indicates that the erupting filament and the overlying loops L3 are heated.
Subsequently, the heated loops L3 are seen sequentially in AIA 94, 335, 211, 193, 171, and 304\,\AA, SUTRI 465\,\AA, and CHASE and NVST H$\alpha$ images. 
This shows the cooling process of the heated loops, which is also supported by the light curves.
Plasmoids formed separately in the filament and the loops L3, and then propagated bidirectionally along them.
New loops L1 (L2) formed, connecting the northern (southern) endpoint of filament and the western (eastern) footpoint of loops L3.
The erupting filament finally disappeared, with the formation of more loops L1 and L2.
The preexisting filament F0 underneath the erupting filament remained stationary, indicating that the filament eruption is a partial eruption of a double-decker filament.

\begin{figure}[ht!]
\centering
\includegraphics[width=0.58\textwidth]{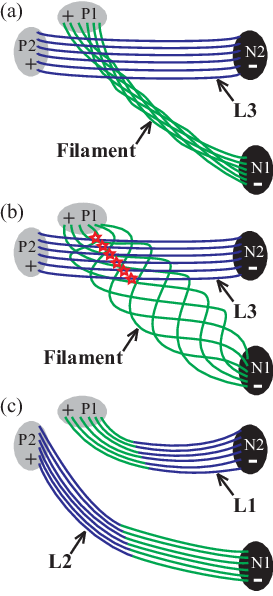}
\caption{Schematic diagrams of the reconnection between the erupting filament and the overlying loops.
The grey (P1 and P2) and black (N1 and N2) ellipses with plus and minus signs represent the positive and negative magnetic fields, respectively.
The green, blue, and green-blue lines separately indicate magnetic field lines of the filament, overlying loops L3, and newly-formed loops L1 and L2.
The red stars in (b) denote the reconnection.
See Section\,\ref{sec:sum} for details.
\label{f:cartoon}}
\end{figure}

According to the AIA, SUTRI, CHASE, and NVST images and the HMI LOS magnetograms, schematic diagrams are provided in Figure\,\ref{f:cartoon} to describe the filament eruption and the reconnection between the erupting filament and the overlying loops L3.
The green (blue) lines in Figure\,\ref{f:cartoon}(a) represent the filament (overlying loops L3), connecting the positive and negative magnetic fields P1 (P2) and N1 (N2).
The filament erupts, showing the untwisting motion, toward the overlying loops L3 (see the green and blue lines in Figure\,\ref{f:cartoon}(b)).
Magnetic reconnection, marked by the red stars in Figure\,\ref{f:cartoon}(b), occurs between the erupting filament and the overlying loops L3.
The newly-reconnected loops L1 and L2 then form (see the green-blue lines in Figure\,\ref{f:cartoon}(c)).

Reconnection between the erupting filament and the overlying loops is observed.
During the eruption, the untwisting motion toward the northern endpoint of the filament changes the field direction of the filament, facilitating the reconnection between the erupting filament and its overlying loops (see Figure\,\ref{f:cartoon}(b)).
The reconnection heats the reconnecting filament and loops.
We hence observe the bright erupting filament and overlying loops L3, with average temperatures of 12.8 and 9.6\,MK, respectively.
The plasmoids, forming and moving along the filament and the loops L3, support that the plasmoid instability takes place during the reconnection \citep{li16a, 2022ApJ...935...85L, 2019A&A...628A...8P}.
After the reconnection, two sets of new loops L1 and L2, separately connecting the erupting filament endpoints and the overlying loop footpoints, formed (see Figure\,\ref{f:cartoon}(c)), and the erupting filament finally disappeared.
This shows the reconfiguration of magnetic structures of the filament and the overlying loops L3 by reconnection \citep{li16a, 2022ApJ...935...85L, 2023ApJ...951L..35C}.
The cooling process of the heated loops L3 is then observed, supported by both the imaging observations and the light curves.
The AIA and NVST light curves increase and peak later than the soft X-ray flux.
This suggests that the enhancement of these light curves originates from the heated and subsequently cooling loops L3, rather than the M4.4 flare emission associated with the failed filament eruption.
The dark material, that appeared and propagated along the overlying loops L3, may represent the condensation of hot plasma in the heated loops \citep{2021ApJ...910...82L, 2021ApJ...919L..21L}, and/or the cooler filament mass moving along the reconnecting and/or newly-reconnected loops. 
It may have an effect on the evolution, e.g., the peak times, of the AIA and NVST light curves of the loops L3. 

Failure of the filament eruption caused by the reconnection between the erupting filament and the overlying loops is reported.
The untwisting motion toward the northern endpoint of the erupting filament shows that the eruption may be triggered by the kink instability \citep{2005ApJ...630L..97T, 2016ApJ...818..148L}. 
Before the filament rapidly decelerated and finally stopped, it rose slowly, and then quickly erupted. 
This initial evolution of the filament eruption is similar to those of the successful eruptions \citep{2020ApJ...894...85C, 2022ApJ...941L...1L, 2024ApJ...967..130L}.
However, the filament failed to escape from the Sun despite that it may exceed the critical height for torus instability \citep{2005ApJ...630L..97T, 2006PhRvL..96y5002K}.
Instead, the failure of the eruption is caused by the reconnection between the erupting filament and the overlying loops \citep{1999ApJ...515L..81A, 2016ApJ...832..106H, 2022ApJ...928..160Z, 2023ApJ...951L..35C, 2023MNRAS.525.5857J}. 
In this study, the erupting filament reconnected with the overlying loops L3, formed two sets of newly-reconnected loops L1 and L2, and finally disappeared.
This indicates that the reconnection ruins the filament completely.
The material and magnetic structures of the filament thus failed to propagate outward into interplanetary space as a CME, forming the failed eruption \citep{2023MNRAS.525.5857J}.
\citet{2019ApJ...877L..28Z} and \citet{2023MNRAS.525.5857J} suggested that the significant rotation of MFR around the vertical axis is important for the failed eruption, as it renders the field direction at the MFR apex almost inverse to the overlying field.
Current sheet, where the reconnection occurs, is hence easily formed between the MFR and the overlying flux.
In this study, the filament rotated mainly around its central axis, rather than the vertical axis, showing the untwisting motion.
Therefore, the untwisting motion toward the northern endpoint of the filament and the expansion of the filament endpoints as well can facilitate the reconnection between the filament and the overlying loops that results in the failed eruption.
Moreover, the smaller (projection) speed (45\,km\,s$^{-1}$) of the filament eruption, and the larger amount of magnetic flux of the overlying loops are also important for the failed eruption.
They provide the overlying loops more time and flux to reconnect with the erupting filament.
More erupting filament is thus ruined by the reconnection, leading to the failed eruption.

\clearpage
\startlongtable
\begin{deluxetable}{cccccc}
\tablecaption{General information of the analyzed data on 2023 September 24 \label{tab:data}}
\tablehead{
\colhead{Telescope} & \colhead{Time} & \colhead{Data Product} & \colhead{Passband} & \colhead{Time Cadence} & \colhead{Spatial Sampling} \\
\colhead{} & \colhead{(UT)} & \colhead{} & \colhead{(\AA)} & \colhead{(s)} & \colhead{(\arcsec\,pixel$^{-1}$)} \\
}
\startdata
SDO/AIA & 00:00-07:00 & Imaging data & 131, 94, 335, 211 & 12 & 0.6 \\
 & & & 193, 171, 304 &  &   \\
\hline
SDO/HMI & 00:00-07:00 & LOS magnetogram & Fe\,{\sc{i}} 6173 & 45 & 0.5 \\
\hline
NVST & 03:11-06:27 & Imaging data & H$\alpha$ 6562.8 & 41.3 & 0.168 \\
\hline
CHASE/HIS & 00:27-00:55 & Imaging data & H$\alpha$ 6562.8 & 71 & 1.05 \\
 & 02:02-02:30 &  &  &  &  \\
 & 03:37-04:05 &  &  &  &  \\
 & 05:12-05:40 &  &  &  &  \\
\hline
SUTRI & 00:02-00:29 & Imaging data & Ne\,{\sc{vii}} 465 & 30 & 1.23 \\
 & 01:10-02:03 &  &  &  &  \\
 & 02:44-02:47 &  &  &  &  \\
 & 03:14-03:37 &  &  &  &  \\
 & 04:18-05:12 &  &  &  &  \\
 & 05:52-06:45 &  &  &  &  \\
\hline
GOES-16 & 00:00-07:00 & Soft X-ray flux & 1-8 & 1 & - \\
\enddata
\end{deluxetable}

\acknowledgments

The authors thank the referee for helpful comments that led to improvements in the manuscript, and thank Drs. Chaowei Jiang, Peng Zou, and Zhe Xu for the discussions. We are indebted to the SDO, NVST, SUTRI, CHASE, and GOES teams for providing the data.
This work is supported by the National Key R\&D Programs of China (2022YFF0503002), the National Natural Science Foundations of China (12073042, U2031109, 12350004, 12273060, and 12273061), the Strategic Priority Research Program (No. XDB 0560000) of CAS, and the Youth Innovation Promotion Association CAS (2023063).
The AIA images are courtesy of NASA/SDO and the AIA, EVE, and HMI science teams.
SUTRI is a collaborative project conducted by the National Astronomical Observatories of CAS, Peking University, Tongji University, Xi'an Institute of Optics and Precision Mechanics of CAS, and the Innovation Academy for Microsatellites of CAS.
The CHASE mission is supported by CNSA.
We acknowledge the usage of Jhelioviewer software \citep{2017A&A...606A..10M} and NASA'S Astrophysics Data System. 


\end{document}